\documentclass[aps,pr,twocolumn,groupadress,showpacs]{revtex4}

\usepackage{graphicx,bm}
\usepackage{amsmath}

\newcommand{\udens}{{\:g/cm$^3$\,}}

\newcommand{\eOCP}{{eOCP}}

\usepackage[colorlinks=true]{hyperref}

\usepackage[ulem=normalem,draft]{changes}
\usepackage{xcolor}
\definecolor{mydarkblue}{rgb}{0.1,0,0.55}
\definecolor{darkgreen}{rgb}{0.1,0.5,0.1}
\definecolor{brown}{rgb}{0.66,0.33,0.0}

\begin{document}

\title{ A Unified Concept of an Effective One Component Plasma  for Hot Dense Plasmas}

\author{Jean Cl\'erouin}
\affiliation{
CEA, DAM, DIF\\
F-91297 Arpajon, France}
\email  {jean.clerouin@cea.fr}

\author{Philippe Arnault}
\affiliation{
CEA, DAM, DIF\\
F-91297 Arpajon, France}

%

\author{Christopher Ticknor}
\affiliation{Theoretical Division, Los Alamos National Laboratory\\
Los Alamos, New Mexico 87545, USA}

\author{Joel D. Kress}
\affiliation{Theoretical Division, Los Alamos National Laboratory\\
Los Alamos, New Mexico 87545, USA}

\author{Lee A. Collins}
\affiliation{Theoretical Division, Los Alamos National Laboratory\\
Los Alamos, New Mexico 87545, USA}

\date{\today}
\begin{abstract}
Orbital-free molecular dynamics simulations are used to benchmark two popular models for hot dense plasmas: the one component plasma (OCP) and the Yukawa model. A unified concept emerges where an effective OCP (eOCP) is constructed from the short-range structure of the plasma. An unambiguous ionization and the screening length can be defined and used for a Yukawa system, which reproduces the long range structure with finite compressibility. Similarly, the dispersion relation of longitudinal waves is consistent with the screened model at vanishing wavenumber but merges with the OCP at high wavenumber. Additionally, the eOCP reproduces the overall relaxation timescales of the correlation functions associated with ionic motion. In the hot dense regime, this unified concept of eOCP can be fruitfully applied to deduce properties such as the equation of state, ionic transport coefficients, and the ion feature in x-ray Thomson scattering experiments.
\end{abstract}

\maketitle

 Matter in the universe is very often found in extreme states, at high pressure ($>$ 1\:Mbar) and high temperature ($>$ 1\:eV).  Such conditions, relevant to planetary interiors \cite{BARA10},  dwarf stars, and  neutron star crusts \cite{DALI09}, can now be reproduced in experiments using high-energy \cite{KONI05} and x-ray free-electron lasers \cite{VINK12} and are routinely met  in inertial confinement fusion studies \cite{LIND04}. This hot dense plasmas (HDP)  regime is an extension to high temperatures ($\simeq $ keV) of the  warm dense matter (WDM) concept \cite{GRAZ14}, more focused on the transition between normal matter and plasmas. In both WDM and HDP regimes, atoms are partially ionized, electrons partially degenerate, and the Coulomb coupling is strong, leading to a liquidlike structure. There is no small parameter enabling a theoretical treatment in perturbation, and the physical description is usually provided by very demanding state-of-the-art quantum \textit{ab initio} simulations. The theoretical description of HDP is a  formidable challenge,  since these methods reach their limits of  applicability. Fortunately, the orbital-free method within a Thomas-Fermi formulation \cite{LAMB06} extends to high temperatures the capability of quantum simulations. It is also desirable to rely on simple models in the first design  and  interpretation of experiments to setup large scale simulations. Such models have to be benchmarked against representative HDP simulations. Here we propose a unified concept of an effective one component plasma that fully describes the complicated nature of strongly correlated plasma without any free parameters. This model offers insights of fundamental focus in plasma physics and is relevant to research areas like astrophysics and fusion science.\\
The one component plasma (OCP) \cite{HANS73,HANS75}  is a popular model which consists of a single species of ions immersed in a neutralizing background of electrons. Its static and dynamical properties depend on only one dimensionless parameter, the Coulomb coupling parameter  $\Gamma = Q^2e^2/ak_BT$  where  $a$ is the Wigner-Seitz (ws) radius $a=(3/4\pi n)^{1/3}$,  $n$ is the ionic density,  $Q$ the ionization, $e$ the fundamental charge, and $T$ the temperature.  Since the OCP model provides a  formulation in which all its properties are either analytical or tabulated, it  is used as a practical representation of Coulomb coupling in many situations encountered in  hot dense plasmas  although it  represents a limiting situation in which the electrons are fully degenerate. 
Attempts to go beyond this simple model belong to the family of screened systems in which the bare coulomb interaction is replaced by a Yukawa potential \cite{HAMA97},  for instance. In the Yukawa model, a screening length is obtained within linear response theory in the small wavenumber $k$ (long distance) limit for given values of  ionization, temperature, and density \cite{STAN15}. In practice, the Yukawa model is deeply modified in the interpretation of x-ray Thomson scattering experiments by the introduction of short-range hard-core corrections that extends further than the first neighbors range \cite{MA14,FLET15}.  All these simplifying assumptions can obscure the diagnostic of the phenomena at play as is revealed by more realistic models \cite{CLER15,PLAG15,STAR15b} and recent experiments \cite{CHAP15}.

These approaches are not satisfactory for actual plasmas  because ionization is not a well-defined quantity and the screening length definition is  somewhat arbitrary. To provide a more realistic  modeling of hot and dense plasmas, we have developed a simple finite temperature Thomas-Fermi  orbital-free formulation coupled with molecular dynamics (OFMD)  \cite{LAMB06}.   With the same inputs as the quantum molecular dynamics simulations with orbitals, {\it i.e.}, atomic number, density, and temperature, the OFMD simulations extend the range of accessible  thermodynamic states without limits on temperature \cite{LAMB06,LAMB06b,CLER13b,KRES11, KRES11b,WHIT14}. A particularly interesting feature is the possibility to perform direct  simulations of mixtures \cite{HORN08,LAMB08,HORN09,KRES10,BURA13,TICK14,SHEP14} to check the validity of  mixing rules for thermodynamical \cite{LAMB08} and transport properties \cite{ARNA13b}. 

In this Letter, we examine the relationship between the  OFMD simulations of plasmas and the simple OCP or Yukawa formulations. We present a unifying concept combining the merits of both formulations. We give arguments supporting the use of the OCP model for the properties involving short-range correlations, including the equation of state \cite{ARNA13} and the transport coefficients. We show that the quantities related to long-range correlations, forming the collective modes, such as the compressibility and the sound speed, are better reproduced by the Yukawa model once the ionization has been consistently defined. We have  investigated two cases of very different atomic numbers relevant to the HDP regime: tungsten twice compressed between 100 and 5000\,eV, and germanium at normal density  between 100 and 800\,eV. We used OFMD in the simplest  formalism (Thomas-Fermi) for simulations. 
Relying on the well-known Thomas-Fermi scaling laws \cite{ARNA13}, we anticipate that our conclusions apply equally to any element in the HDP regime.
\begin{figure}[!t]
\begin{center}
\includegraphics[scale=0.44]{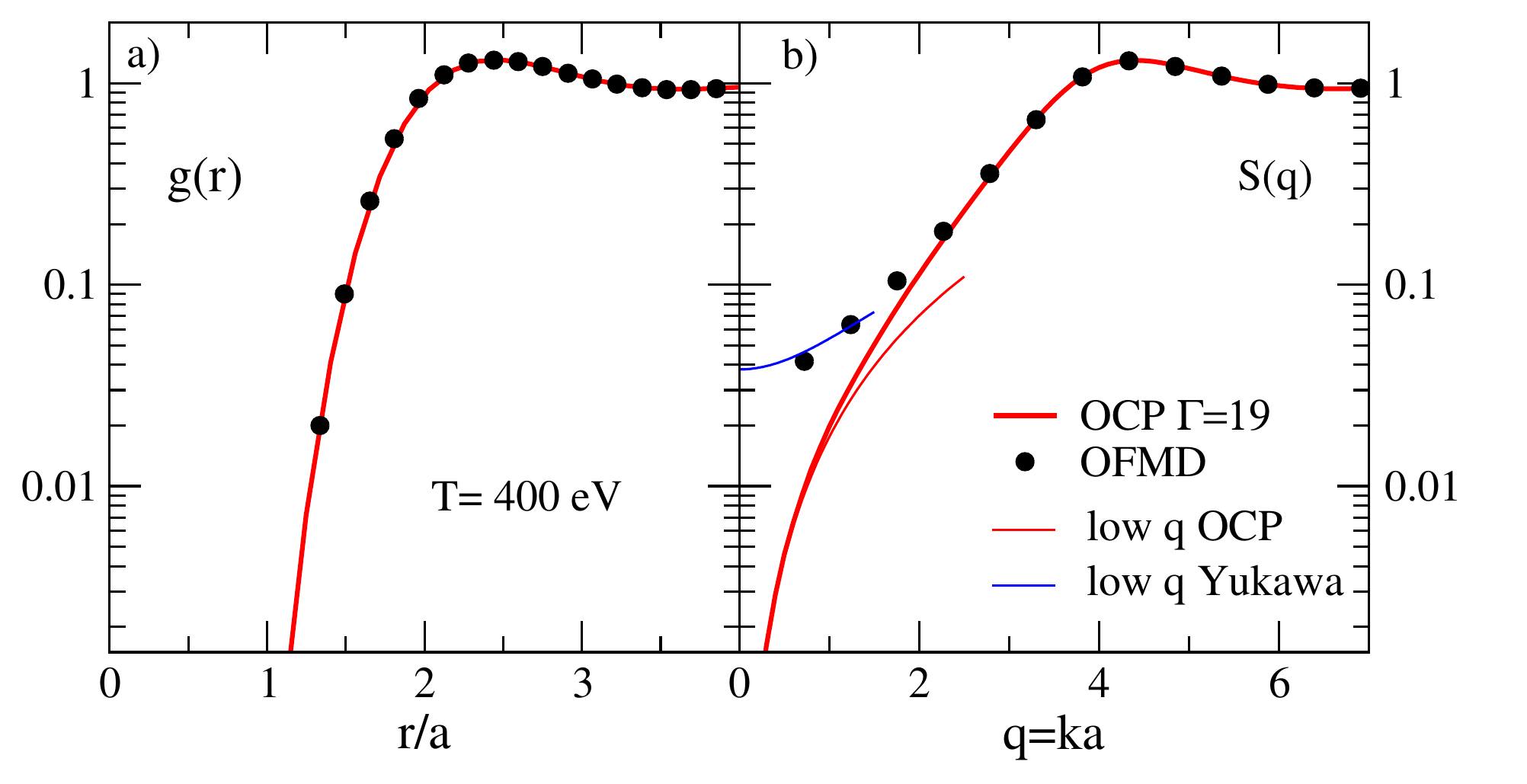}
\caption{a) OFMD pair distribution function for tungsten at 40\udens and 400\,eV (black  points) compared with the OCP result (red solid line) at $\Gamma=19$. b) Corresponding static structure factor. The blue line is the low $q$ expansion of Yukawa structure factor using the effective  ionization of the OCP fit. The thin red line is the OCP  low-$q$ expansion. Note the logarithmic scale to emphasize the differences at low wavenumber $q$.}
\label{gdr_400eV}
\end{center}
\end{figure}

An interesting feature, evidenced in \cite{CLER13b}, is that  the structure of the plasma, revealed by the pair  distribution function (pdf) generated from OFMD simulations,  can be precisely fitted by the  OCP (see also \cite{RECO09}). This procedure defines the effective OCP (\eOCP) with the effective coupling parameter $\Gamma_e=Q_e^2e^2/ak_BT$ and  ionization $Q_e$. A similar procedure has been also invoked by Ott \textit{et al.} \cite{OTT14} to characterize the coupling for Yukawa systems, which gives results very close  to an adjustment {\it by hand}. Ott's method provides a quantitative basis to the effective OCP concept \cite{SUPP1}.    We show in Fig.\:\ref{gdr_400eV}a  such an adjustment extracted from a series of simulations on tungsten at 40\udens and between  100 and 5000\:eV. We chose a temperature of 400\:eV which is just in the region of the $\Gamma$-plateau where the structure is independent of the temperature   \cite{ARNA13,CLER13}. This peculiar evolution is due to the increase of ionization that compensates for the increase of temperature.  It is worth noting that the structure is exactly the same with  exchange (TFD) and gradient-corrected functionals \cite{PERR79}, leading to the same effective coupling. Values of plasma parameters deduced from the \eOCP\,  analysis are given in Table \ref{Csts} for tungsten at 40\udens  between 100  and 5000\,eV, and germanium at 5\udens between 100 and 800\,eV. The details of the OFMD simulations and various formulas can be found in Supplemental  Material \,\cite{SUPP1}.

\begin{table}[t]
\begin{center}
\begin{tabular}{cccccccc}
\hline
Elt	& T		& $\Gamma_e$		&Q$_e$	&	Q$_{TF}$	&$\theta$	& $\kappa$&$T_{\omega_p}$	\\
	& eV		&				&		&			&		&		 &	a.u.			\\	
\hline\
	&100		&19			&12.7		&14.2		&1.9		&	2.1	&	580				\\	
W	&200		&19			&18.0		&19.5		&3.0		&	1.8	&	409			\\
	&400		&19			&25.4		&27.9		&4.7		&	1.5	&	290			\\	
	&800		&19			&35.9		&39.4		&7.5		&	1.3	&	205		\\
	&1200	&17			&41.6		&46.9		&10.		&	1.1	&	177		\\
	&5000	&10			&65.1		&67.3		&31.		&	0.7	&	113		\\

\hline
	&100		&8			&10.0		&10.7		&4.7		&	1.6	&	822			\\	
Ge	&200		&8			&14.1		&15.4		&7.5		&	1.3	&	583			\\
	&400		&8			&20.0		&20.9		&12.		&	1.1	&	411			\\		
	&800		&7			&26.4		&26.0		&20.	&		0.9	&	312			\\
	
\hline
\end{tabular}
\end{center}
\caption{Ionizations and plasma parameters for a tungsten plasma at 40\udens\, and a germanium plasma at 5.3\udens. $\theta = k_BT/E_F$ where $E_F$ is the Fermi energy. $\kappa$ is the inverse screening length at finite temperature in units of the ws radius $a$. Plasma   periods $T_{\omega_p}=2\pi/ \omega_p$ are given in atomic units.}
\label{Csts}
\end{table}

We see in Fig.\:\ref{gdr_400eV}a for tungsten at 400\,eV  that the  \eOCP\, pdf at $\Gamma_e=19$ perfectly matches  the pdf obtained from OFMD. From the value of the coupling parameter $\Gamma_e$, we can deduce an effective ionization $Q_e=25.4$ which appears to be  10\% lower than an estimate  within the average atom (AA) framework using the same Thomas-Fermi functional, $Q_{TF}$ (see Table \ref{Csts}).   This suggests that the piling up of electrons around each ion is different in the OFMD and AA approaches, leading to different ionization and screening at short distance \cite{MURI13}. In any case, both approaches here account for the \emph{nonlinear} contributions to screening close to the ions, contrary to the Yukawa model where screening is always considered within linear response. Within the \eOCP\,  model, the nonlinear screening \textit{at short distance} is embodied in the effective charge $Q_e$.   The good agreement between the \eOCP\,  and the  OFMD results at short distance  deteriorates at long distance (small $q=ka$) as revealed by the calculation of the static structure factor $S(q)$ shown in Fig.\,\ref{gdr_400eV}b.  At vanishing $q$,  $S_{\mathrm{OCP}}(q)$ goes to zero as  $q^2 / 3 \Gamma$ \cite{BAUS80} due to the long range of the Coulomb potential, whereas  $S_{\mathrm{OFMD}}(q)$ goes to a finite value proportional to the isothermal compressibility. Actually, screening effects must be introduced at long distance. Assuming a Yukawa pair potential with an inverse finite temperature screening length $\kappa=k_{FT} a$ \cite{DHAR81,STAN15,Yukacomment},  the resulting $S_\mathrm{Y}(q)$ tends to a finite value as $(q^2  + \kappa^2) / (q^2 + \kappa^2+ 3 \Gamma)$ at vanishing $q$.  Using the effective charge $Q_e$ as a definition of the ionization to compute the  screening constant $\kappa$, the low $q$ expansion  of $S_\mathrm{Y}(q)$ connects seamlessly with  the OFMD results. For tungsten at 400\,eV, the low $q$ expansion of  $S_\mathrm{Y}(q)$ is given as a blue line on Fig.\:\ref{gdr_400eV}b. An extensive comparison with OFMD results will be presented in a forthcoming paper. This connection between OCP and Yukawa models through the definition of an effective charge is absent in traditional modeling where the ionization used to compute the screening length is left as a free parameter. The ionization is often assumed to be complete or deduced from an average atom calculation. Here we extract the effective charge from the static structure of the pdf. It can also be parameterized from a limited set of simulations using the Thomas-Fermi scaling laws. We left the presentation of this parameterization to a future paper.

A straightforward application of the eOCP concept concerns the equation of state. Very often, the ion thermal part is  difficult  to evaluate and is simplified or taken as an interpolation between the solid and  the perfect gas. In the OFMD simulations this contribution is explicitly computed.  In the eOCP approach the ion thermal contribution is constructed from analytical OCP fits \cite{SLAT80} taken at $\Gamma_e$ and the electron contribution  from the corresponding finite temperature Fermi gas, as fitted by Nikiforov {\it et al.} \cite{NIKI05}, at the electronic density corresponding to $Q_e$. We show in Table \ref{TablePress} for the case of germanium that the sum of these two contributions  $P_{\mathrm{eff}}$ agree to better than 10\% with the SESAME equations of state \cite{SESA92,ARNA13} or the present direct simulations with  OFMD.

\begin{table}[t]
\begin{center}
\begin{tabular}{cccccc}
\hline
Elt	& T		& $\Gamma_e$		&P$_{\mathrm{OFMD}}$	&P$_{\mathrm{eff}}$	&P$_{\mathrm{SESA}}$	\\
	& eV		&				&  Mbar		&Mbar		& Mbar		\\	
\hline\
	&100		&8			&85			&80		&80		\\	
Ge	&200		&8			&229			&217		&221		\\
	&400		&8			&608			&596		&596		\\		
	&800		&7			&1509		&1551	&1494		\\
	
\hline
\end{tabular}
\end{center}
\caption{Equation of state of germanium at $5.3$\udens.  P$_{\mathrm{OFMD}}$ is the pressure obtained by simulations, P$_{\mathrm{eff}}$ is the sum of the eOCP contribution and the electronic component as given by Nikiforov \cite{NIKI05} (see Supplemental Material \cite{SUPP1}) and P$_{\mathrm{SESA}}$ is the corresponding SESAME equation of state \cite{SESA92}.}
\label{TablePress}
\end{table}
%
\begin{figure}[!t]
\begin{center}
\includegraphics[scale=0.55]{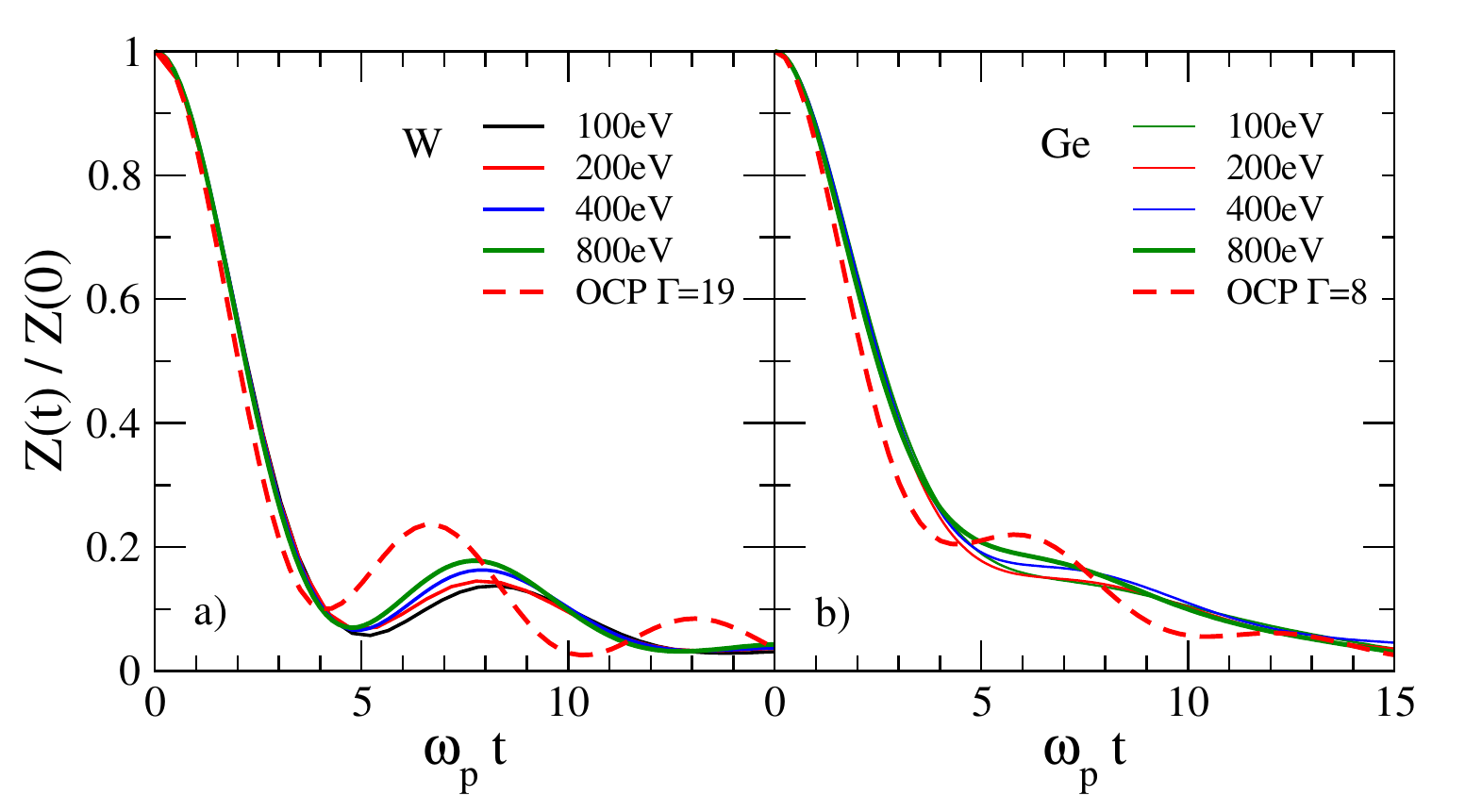}
\caption{ a) OFMD velocity autocorrelation functions of tungsten at 40\udens as a function of time in units of the  inverse plasma frequency for each temperature, given in Table\,\ref{Csts}, compared with the effective OCP  one (red dashed line).  b) Same as a) except for germanium at 5.3\udens.}
\label{DoubleZ_W}
\end{center}
\end{figure}

We turn now to the dynamical properties. It is well-known that the long-wavelength excitations of the charged versus neutral systems are notably different. Baus \cite{BAUS75} described these differences using a kinetic theory of the fluctuation spectra. The long range behavior of the Coulomb potential, with its singularity at $q \to 0$, is responsible for the various differences, here addressed using OFMD simulations. 
 First, we consider the velocity autocorrelation function $Z(t)$ (VACF), which characterizes the individual motion and coupling with the collective modes. As such, it depends on correlation at both short and long ranges. 
Fig.\,\ref{DoubleZ_W}a  shows the VACF of tungsten between 100 and 800\:eV  in  units of the inverse effective plasma frequency $\omega_{p}^2=4\pi n Q_{e}^2e^2/ M$ of each case, which is ionization dependent ($M$ is the ion mass). We observe  that all VACFs are almost synchronized over a wide range of temperature, which reflects the $\Gamma$-plateau behavior.  Notice that the short time behavior  stays close to the \eOCP.   This indicates that the corresponding Einstein frequencies $\omega_E$  are close to the OCP values  of $\omega_p/\sqrt{3}$.  The relaxation timescales of the VACFs of \eOCP\,  and  OFMD are comparable although the frequencies of oscillations around the average are different.  The same behavior is also observed  in Fig.\,\ref{DoubleZ_W}b for germanium with weaker oscillations corresponding to a lower effective coupling parameter $\Gamma_e$.
It is possible to get better agreement with the eOCP VACFs by a renormalization of the eOCP mass that depends on screening. This is beyond the scope of this paper and will be treated in a forthcoming paper.

\begin{figure}[!ht]
\begin{center}
\includegraphics[scale=0.55]{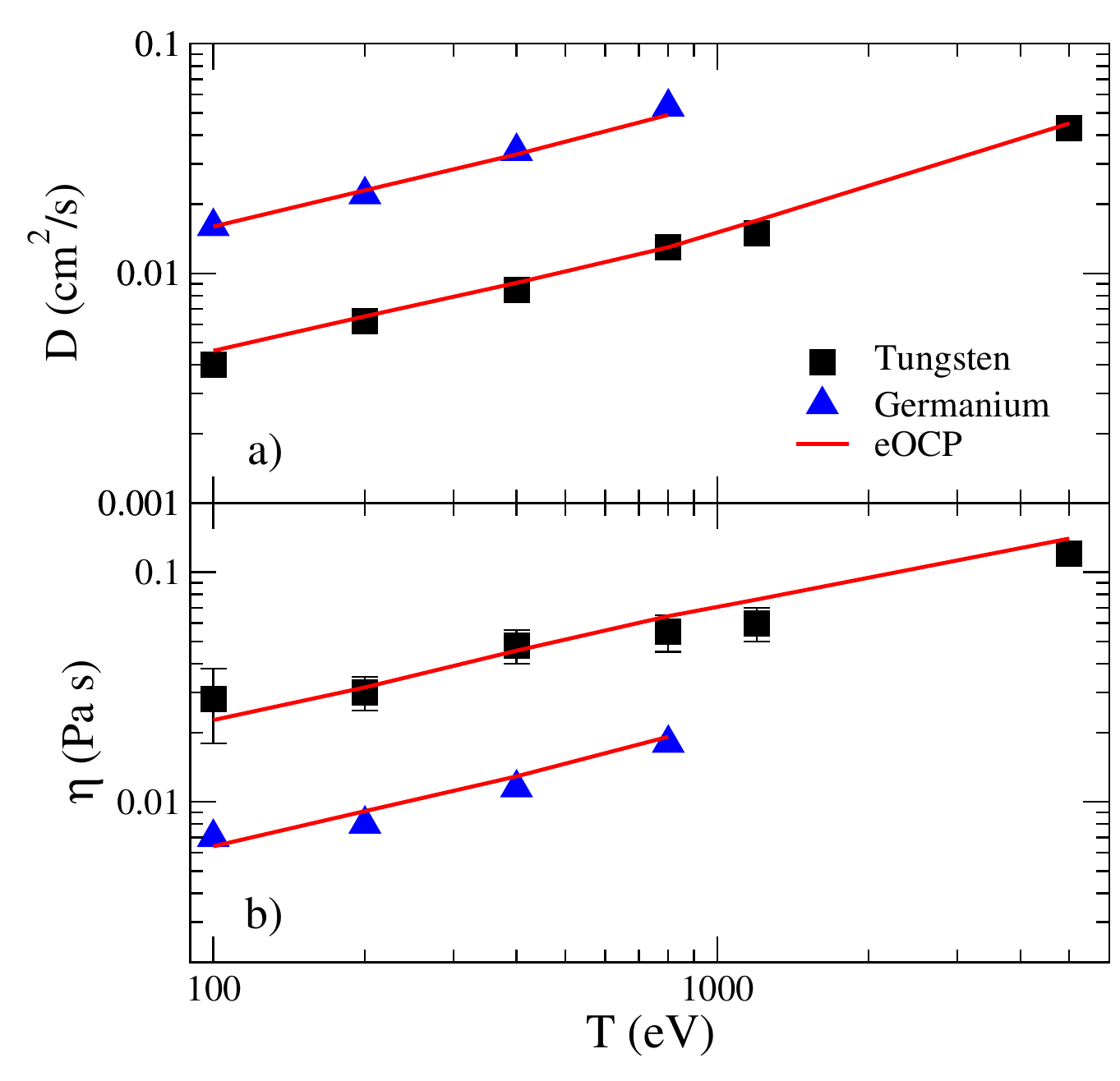}
\caption{Comparison between OFMD simulations (black squares for tungsten at 40\udens\, and blue triangles for germanium at 5.3\udens) and effective OCP (red solid lines) for (a) diffusion and (b) viscosity. }
\label{trans}
\end{center}
\end{figure}

 The preceding analysis suggests that  the \eOCP\, concept can be used to predict transport coefficients by using standard OCP fits (see \cite{ARNA13b} and references therein) with the effective coupling parameter $\Gamma_e$. Both diffusion coefficients and viscosities are obtained from OFMD simulations by the Green-Kubo relations (see \cite{BAST05,DALI06,DALI09b} for diffusion and \cite{MEYE14} for viscosity). Good agreement for viscosity and diffusion for both tungsten and germanium  is found with the \eOCP\, formulation as shown in   Fig.\,\ref{trans}.   Comparisons for plasmas of other species for such an approach using the TF ionization can be found in Ref.\,\cite{ARNA13b}.

\begin{figure}[!t]
\begin{center}
\includegraphics[scale=0.5]{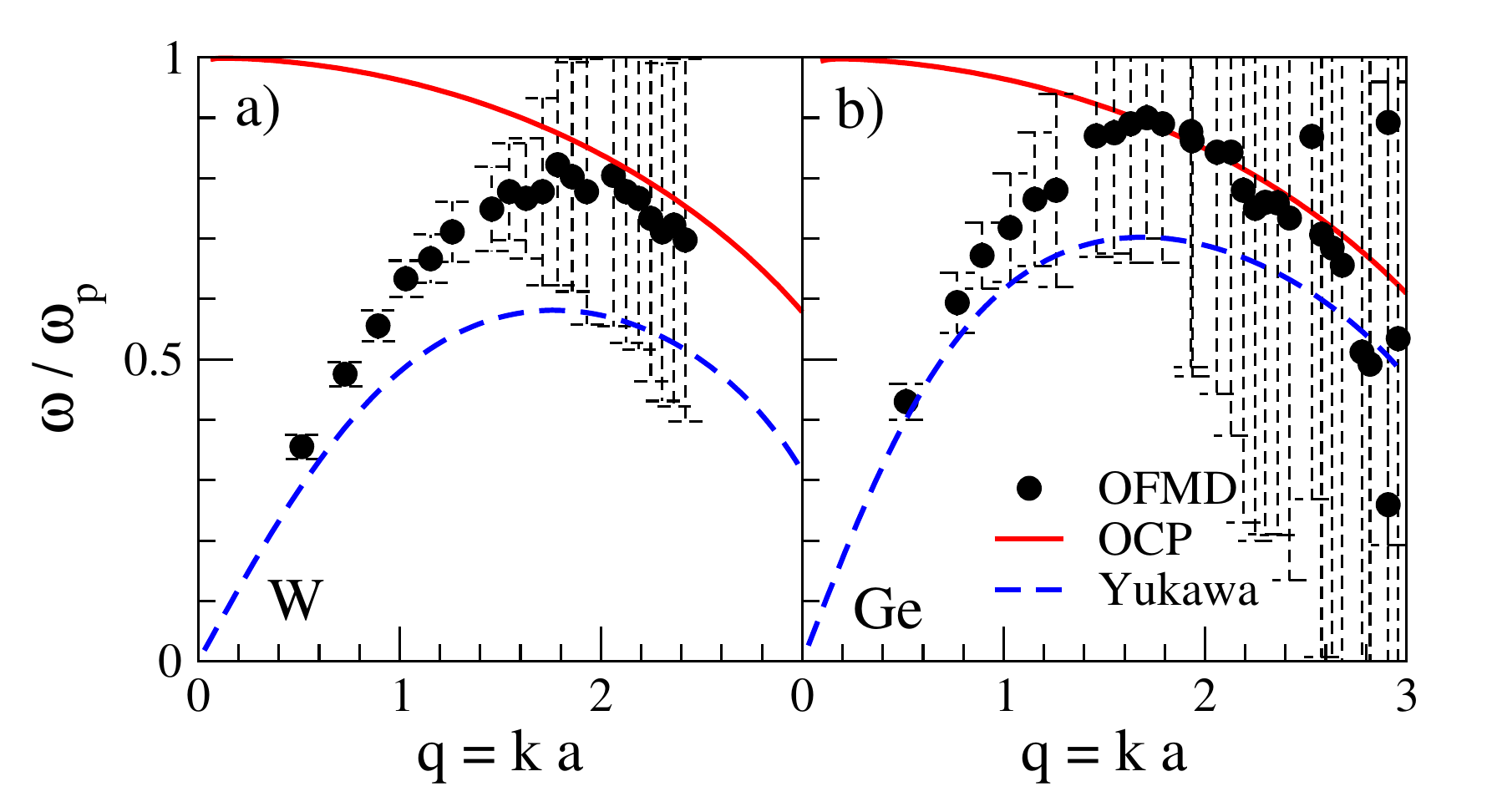}
\caption{Dispersion relations of longitudinal waves for a) tungsten at 40\udens and 400\,eV and b) germanium at 5.3\udens and 400\,eV. Black points: OFMD peak value with FWHM. The dashed blue line is the low $q$ dispersion relation for Yukawa system in the quasi-localized charge approximation proposed by Rosenberg and Kalman \cite{ROSE97}, using the equation of state of Ref.\,\cite{HAMA97}. The red line is the OCP dispersion relation at $\Gamma=19$ for tungsten, and $\Gamma=8$ for germanium, respectively.}
\label{Disp_400eV}
\end{center}
\end{figure}

%

Finally, collective modes are evidenced by the dynamical structure factor $S(q,\omega)$ \cite{HANS06}, which is of particular importance for x-ray scattering experiments. Here we focus on ion collective properties. The calculation of this quantity is well documented, and we follow White {\it et al.} \cite{WHIT13}  and R\"uter {\it et al.} \cite{RUTE14}. By collecting peak frequencies of the OFMD simulations  of $S(q,\omega)$ versus $q$, and full widths at half maximum (FWHMs) of these features, we produce the dispersion relations shown in Fig.\,\ref{Disp_400eV}a for tungsten and Fig.\,\ref{Disp_400eV}b for germanium, which can be fitted by $\omega=c_{s} q/a$ at low $q$, yielding the sound speed $c_s$.  As in the case of the static structure, we observe good agreement with the Yukawa dispersion relation at vanishing $q$. 
We used the relation proposed by Rosenberg and Kalman \cite{ROSE97} within the quasi-localized charge approximation, which is particularly well adapted to the wave dispersion in strong-coupling situations \cite{DIAW15}. For finite wavenumber (typically $q>0.5$), the frequencies of the  OFMD modes are slowly drifting out of the Yukawa curve and join smoothly with the \eOCP\, values for $q>1.5$.

%
To summarize, a unified concept for hot dense plasmas combining the OCP and Yukawa models is proposed. Its merits have been assessed using orbital-free molecular dynamics simulations in the hot and dense regime. The OCP and Yukawa models give complementary information about the simulated plasmas, providing a comprehensive description of their static and dynamical properties. The concept of an effective OCP connects these models through an effective ionization that is unambiguously defined. The eOCP facet is well adapted for short-range correlations  and for a straightforward evaluation of the equation of state and transport coefficients. The properties related to the correlations at large distance, like  the sound speed and the compressibility, need an explicit account of the electron screening. Here the Yukawa facet of this unified concept, based on an eOCP ionization, is a sensible approximation in this range where linear response theory applies.

This work has been performed under the NNSA/DAM collaborative agreement P184. We specially thank Flavien Lambert for providing his OFMD code. PA would like to thank Nicolas Desbiens for fruitful discussions. The Los Alamos National Laboratory is operated by Los Alamos National Security, LLC for the National Nuclear Security Administration of  the U.S. Department of Energy under Contract No. DE-AC52-06NA25396.
%

\end{document}


\title{ ``A Unified Concept of an Effective One Component Plasma  for Hot Dense Plasmas'' \\
--- Supplemental Material --- }

\author{Jean Cl\'erouin}
\affiliation{
CEA, DAM, DIF\\
F-91297 Arpajon, France}
\email  {jean.clerouin@cea.fr}

\author{Philippe Arnault}
\affiliation{
CEA, DAM, DIF\\
F-91297 Arpajon, France}

%

\author{Christopher Ticknor}
\affiliation{Theoretical Division, Los Alamos National Laboratory\\
Los Alamos, New Mexico 87545, USA}

\author{Joel D. Kress}
\affiliation{Theoretical Division, Los Alamos National Laboratory\\
Los Alamos, New Mexico 87545, USA}

\author{Lee A. Collins}
\affiliation{Theoretical Division, Los Alamos National Laboratory\\
Los Alamos, New Mexico 87545, USA}
\maketitle

\section{Simulations}
We have used the orbital-free molecular dynamics package (OFMD)  to simulate a collection of 432  nuclei in a 3-dimensional box with the Thomas-Fermi functional. We tested Thomas-Fermi-Dirac and gradient  corrected functionals \cite{PERR79} with very little differences in the structure and pressure in this range of temperature, as previously mentioned \cite{WHIT14}.
The divergence of the electron-nucleus potential was regularized at each thermodynamical condition. The cutoff radius was chosen as 30\% of the Wigner-Seitz radius, sufficient to prevent overlap of the regularization spheres. The number of plane waves describing the local electronic density was then adjusted to converge the thermodynamic properties to within 1\%. 
 The time-step was adapted to each temperature to follow the corresponding increase of the plasma frequency  $\omega_p^2=4\pi Q_e^2n^{2} e^{2}/M$, where $M$ and $n$ are, respectively,  the ionic mass and density. This characteristic frequency depends on the ionization $Q_{e}$.  To ensure good energy conservation, $\Delta t$ has been varied from 50 atomic units (au) at 100\,eV to 10\,au at 5000\,eV. We simulate between 5000 and 10000 time steps, rejecting the first 1000 time steps. Thermodynamical, structural, and transport properties were obtained for each simulated state.

\section{Formulations}
\subsection{Effective coupling}
\begin{figure}[!t]
\begin{center}
\includegraphics[scale=0.4]{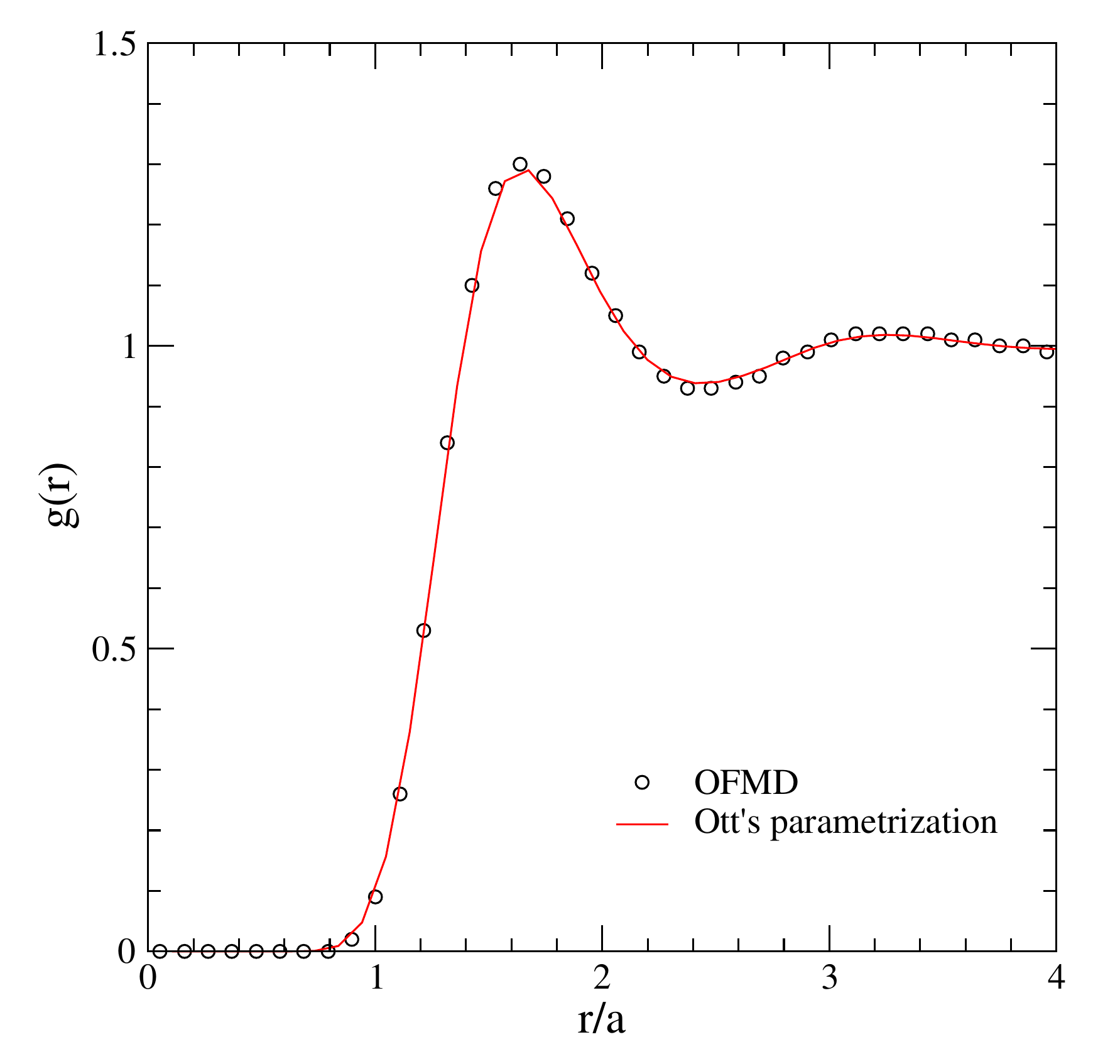}
\caption{ OFMD pair distribution function for tungsten at 40\udens and 400\,eV (black \deleted{solid} points) compared with the OCP result (red solid line) at $\Gamma$=18.2, predicted by the Ott's model \protect\cite{OTT14} with $g_{max}=1.3$ and $r_{1/2}=1.20$ .}
\label{gdr_400eV}
\end{center}
\end{figure}

The ion coupling parameter $\Gamma=Q^2e^2/ak_BT$ uses the mean ionic radius  $a=(3/4\pi n_i)^{1/3}$. All distances or wave numbers are expressed in the $a$  unit. The effective coupling parameter $\Gamma_e$ is obtained by the comparison of the OCP pair distribution function (pdf) with the OFMD simulations. The optimization can be performed {\it by hand}, using tabulated OCP pdfs \cite{ROGE83}  or by using the Ott {\it et al.} \cite{OTT14} method, which requires the first distance $r_{1/2}$ for which $g(r)=0.5$, in units of $a$, at low coupling or the intensity of the first maximum $g_{max}$ at high coupling. We use the Ott parametrization at zero screening to get the effective OCP at low coupling 
 \begin{eqnarray}
\Gamma_e & = & 1.238\exp{(1.575 r_{1/2}^3)}-0.931\\
&&r_{1/2} < 1.3, \nonumber
\end{eqnarray}
and  at high coupling
 \begin{eqnarray}
\Gamma_e & = & 22.40-70.09 g_{max}+52.60g_{max}^2 \\
&&1.4  <   g_{max} < 2.4.  \nonumber
\end{eqnarray}
This defines an effective OCP within 5\% agreement on the coupling parameter. An example of the adjustment predicted by the Ott model is shown in Fig.\,\ref{gdr_400eV}.

 The effective charge $Q_{e}$ is obtained by
\begin{equation}
Q_{e}=\sqrt{\Gamma_{e} a k_{B}T}/e,
\end{equation}
which defines a mean electronic distance $a_e=(3/4\pi n_e)^{1/3}=a/Q_e^{1/3}$.
We use a definition of the Thomas-Fermi screening length at finite temperature \cite{DHAR81} that we fit by 
\begin{equation}
\kappa_{FT}=\sqrt{   {\frac{1-0.69\theta(1-\exp{(-2.26\theta)})+{2 \over 3}\theta^{4}} {1+\theta^5} } }  \kappa_0,
\end{equation}
where $\theta=T/T_F$ is the usual degeneracy parameter ($k_BT_F=(\hbar^2/2 m_e) ( 3\pi^2 Q_e n_i )^{2/3}$) and $\kappa_0$ is the usual zero-temperature screening length $\kappa_0=2(k_F/ \pi a_B)^{1/2}$with $a_B$ the Bohr  radius.
In units of $a$ and using the effective ionization $Q_e$, we get $$\kappa_0 a= \left ( 12 Q_e/\pi \right )^{1/3} \sqrt{a_e/a_B}=\left ( 12/\pi \right )^{1/3} Q_e^{1/6} \sqrt{a/a_B}.$$ 
%
%
\subsection{Equation of state}
%
The pressure is given by the effective pressure $P_{\mathrm {eff}}$, 
\begin{subequations}
\label{pgam}
\begin{equation}
P_{\mathrm{eff}}=P_{\mathrm{OCP}}+P_{\mathrm{ele}}.
\end{equation}
For the ionic OCP part, we used the fit given  by Slattery {\it et al} \cite{SLAT80} without the Madelung contribution
\begin{equation}
   P_{\mathrm{OCP}}/nk_B T  = 1 + {1 \over 3} \left [  b\Gamma^{1/4} + c\Gamma^{-1/4}+d \right ],
\end{equation}
with   b  =  0.94544, c  =  0.17954 and   d  =  -0.80049.\\
For the electronic component, we use an interpolation formula between the Fermi gas and the perfect gas due to Nikiforov {\it et al} \cite{NIKI05}, with the effective ionization $n_e=Q_e n$,
\begin{equation}
   P_{\mathrm{ele}}/n_e  =  \left [ ( k_BT)^3 + 3.36n_e(k_BT)^{3/2} +{\frac {9\pi^4}{125}} n_e^2 \right ]^{1/3},
   \label{niki}
\end{equation}
\end{subequations}
where  atomic units are used (1 a.u. of pressure =294 Mbar).

\subsection{Transport coefficients}
For the diffusion, we used Daligault's fit \cite{DALI06,DALI09b} 
\begin{equation}
D/D_0=\sum_{i=0}^3 a_i \Gamma^i/ \sum_{i=0}^3 b_i\Gamma^i, 
\end{equation}
with  $D_0=\omega_pa^2$ and the set of coefficients corresponding to $\Gamma > 2$  ( see Table \ref{default}).
\begin{table}[ht]
\begin{center}
\begin{tabular}{cccccccc}
a$_0$	&a$_1$	&10$^3$\,a$_2$	&10$^5$ a$_3$	&b$_0$	&b$_1$	&b$_2$	&10$^3$\,b$_3$\\
\hline \hline
59.74	&30.10	&1.37	&-2.403			&-32.11	&56.25	&1.241	&3.72\\
\hline
\end{tabular}
\end{center}
\caption{Coefficients for the diffusion fit.}
\label{default}
\end{table}

For the viscosity we used Bastea's fit \cite{BAST05}
\begin{equation}
{\frac{\eta}{\eta_0}}=A \Gamma^{-2}+B \Gamma^{-s}+C \Gamma,
\end{equation}
with $A$=0.482, $B$=0.629 and $C$=1.88\:10$^{-3}$ and $s$=0.878.
$\eta_{0}=n M a^2 \omega_p$ is the natural unit for viscosity.
%
\subsection{Wave dispersion}
%
%
The wave dispersion relation for the longitudinal modes is computed within the quasilocalized charge approximation of Rosenberg and Kalman \cite{ROSE97}. In the long-wavelength limit it is given by 
\begin{equation}
{\frac{\omega^2}{\omega_p^2}}={\frac{q^2}{q^2+\kappa^2}}+{\frac{q^2}{\Gamma}}\left [  {\frac{4}{45}} u_c   -{\frac{2}{45}} y  {\frac{\partial u_c}{\partial y}} +{\frac{4}{15}} y^2  {\frac{\partial^2 u_c}{\partial y^2}}                     \right ],
\end{equation}
where $u_c=U_c/Nk_{B}T$ is the normalized correlation energy, $y=\kappa^2$.
 Hamaguchi et al. \cite{HAMA96} gave a fit for Yukawa systems
 \begin{equation}
u_c(\kappa,\Gamma)=a(\kappa)\Gamma+b(\kappa)\Gamma^s+c(\kappa) +d(\kappa)\Gamma^{-s},
\end{equation}
with $s=1/3$ and the following variations for coefficients
\begin{eqnarray}
a(\kappa)		&=&E_{bcc}+\delta a(\kappa)\\								\nonumber
E_{bcc}(\kappa)&=&-0.895929	-0.103731\kappa^2		+0.003084\kappa^4\\ 	\nonumber
			&&	-0.000131\kappa^6\\	\nonumber
\delta a(\kappa)	&=&-0.003366	+0.000660 \kappa^2 		-0.000089 \kappa^4\\ 	\nonumber
b(\kappa)		&=&+0.565004	-0.026134  \kappa^2		-0.002689 \kappa^4\\	\nonumber
c(\kappa)		&=&-0.206893	-0.086384  \kappa^2		+0.018278\kappa^4\\		\nonumber
d(\kappa)		&=&-0.031402	+0.042429  \kappa^2		-0.008037\kappa^4
\end{eqnarray}
These coefficients are valid at weak screening ($\kappa < 1$) but we extended their range by interpolating with results at $\kappa=1.2$ and $\kappa=1.5$ \cite{HAMA97}.
